\documentclass[11pt]{article}
\usepackage{amssymb}

\def\PsfigVersion{1.9}
\ifx\undefined\psfig\else\endinput\fi

\let\LaTeXAtSign=\@
\let\@=\relax
\edef\psfigRestoreAt{\catcode`\@=\number\catcode`@\relax}
\catcode`\@=11\relax
\newwrite\@unused
\def\ps@typeout#1{{\let\protect\string\immediate\write\@unused{#1}}}
\ps@typeout{psfig/tex \PsfigVersion}

\def\figurepath{./}

\def\@nnil{\@nil}
\def\@empty{}
\def\@psdonoop#1\@@#2#3{}
\def\@psdo#1:=#2\do#3{\edef\@psdotmp{#2}\ifx\@psdotmp\@empty \else
    \expandafter\@psdoloop#2,\@nil,\@nil\@@#1{#3}\fi}
\def\@psdoloop#1,#2,#3\@@#4#5{\def#4{#1}\ifx #4\@nnil \else
       #5\def#4{#2}\ifx #4\@nnil \else#5\@ipsdoloop #3\@@#4{#5}\fi\fi}
\def\@ipsdoloop#1,#2\@@#3#4{\def#3{#1}\ifx #3\@nnil
       \let\@nextwhile=\@psdonoop \else
      #4\relax\let\@nextwhile=\@ipsdoloop\fi\@nextwhile#2\@@#3{#4}}
\def\@tpsdo#1:=#2\do#3{\xdef\@psdotmp{#2}\ifx\@psdotmp\@empty \else
    \@tpsdoloop#2\@nil\@nil\@@#1{#3}\fi}
\def\@tpsdoloop#1#2\@@#3#4{\def#3{#1}\ifx #3\@nnil
       \let\@nextwhile=\@psdonoop \else
      #4\relax\let\@nextwhile=\@tpsdoloop\fi\@nextwhile#2\@@#3{#4}}
\ifx\undefined\fbox
\newdimen\fboxrule
\newdimen\fboxsep
\newdimen\ps@tempdima
\newbox\ps@tempboxa
\long\def\fbox#1{\leavevmode\setbox\ps@tempboxa\hbox{#1}\ps@tempdima\fboxrule
    \advance\ps@tempdima \fboxsep \advance\ps@tempdima \dp\ps@tempboxa
   \hbox{\lower \ps@tempdima\hbox
  {\vbox{\hrule height \fboxrule
          \hbox{\vrule width \fboxrule \hskip\fboxsep
          \vbox{\vskip\fboxsep \box\ps@tempboxa\vskip\fboxsep}\hskip
                 \fboxsep\vrule width \fboxrule}
                 \hrule height \fboxrule}}}}
\fi
\newread\ps@stream
\newif\ifnot@eof       
\newif\if@noisy        
\newif\if@atend        
\newif\if@psfile       
{\catcode`\%=12\global\gdef\epsf@start{
\def\epsf@PS{PS}
\def\epsf@getbb#1{%
\openin\ps@stream=#1 \ifeof\ps@stream\ps@typeout{Error, File #1 not
found}\else
   {\not@eoftrue \chardef\other=12
    \def\do##1{\catcode`##1=\other}\dospecials \catcode`\ =10
    \loop
       \if@psfile
      \read\ps@stream to \epsf@fileline
       \else{
      \obeyspaces
          \read\ps@stream to \epsf@tmp\global\let\epsf@fileline\epsf@tmp}
       \fi
       \ifeof\ps@stream\not@eoffalse\else
       \if@psfile\else
       \expandafter\epsf@test\epsf@fileline:. \\%
       \fi
          \expandafter\epsf@aux\epsf@fileline:. \\%
       \fi
   \ifnot@eof\repeat
   }\closein\ps@stream\fi}%
\long\def\epsf@test#1#2#3:#4\\{\def\epsf@testit{#1#2}
            \ifx\epsf@testit\epsf@start\else
\ps@typeout{Warning! File does not start with `\epsf@start'.  It may
not be a PostScript file.}
            \fi
            \@psfiletrue} 
{\catcode`\%=12\global\let\epsf@percent=
\long\def\epsf@aux#1#2:#3\\{\ifx#1\epsf@percent
   \def\epsf@testit{#2}\ifx\epsf@testit\epsf@bblit
    \@atendfalse
        \epsf@atend #3 . \\%
    \if@atend
       \if@verbose{
        \ps@typeout{psfig: found `(atend)'; continuing search}
       }\fi
        \else
        \epsf@grab #3 . . . \\%
        \not@eoffalse
        \global\no@bbfalse
        \fi
   \fi\fi}%
\def\epsf@grab #1 #2 #3 #4 #5\\{%
   \global\def\epsf@llx{#1}\ifx\epsf@llx\empty
      \epsf@grab #2 #3 #4 #5 .\\\else
   \global\def\epsf@lly{#2}%
   \global\def\epsf@urx{#3}\global\def\epsf@ury{#4}\fi}%
\def\epsf@atendlit{(atend)}
\def\epsf@atend #1 #2 #3\\{%
   \def\epsf@tmp{#1}\ifx\epsf@tmp\empty
      \epsf@atend #2 #3 .\\\else
   \ifx\epsf@tmp\epsf@atendlit\@atendtrue\fi\fi}

\chardef\psletter = 11 
\chardef\other = 12

\newif \ifdebug 
\newif\ifc@mpute 
\c@mputetrue 

\let\then = \relax
\def\r@dian{pt }
\let\r@dians = \r@dian
\let\dimensionless@nit = \r@dian
\let\dimensionless@nits = \dimensionless@nit
\def\internal@nit{sp }
\let\internal@nits = \internal@nit
\newif\ifstillc@nverging
\def \Mess@ge #1{\ifdebug \then \message {#1} \fi}

{ 
    \catcode `\@ = \psletter
    \gdef \nodimen {\expandafter \n@dimen \the \dimen}
    \gdef \term #1 #2 #3%
           {\edef \t@ {\the #1}
        \edef \t@@ {\expandafter \n@dimen \the #2\r@dian}%
        \t@rm {\t@} {\t@@} {#3}%
           }
    \gdef \t@rm #1 #2 #3%
           {{%
        \count 0 = 0
        \dimen 0 = 1 \dimensionless@nit
        \dimen 2 = #2\relax
        \Mess@ge {Calculating term #1 of \nodimen 2}%
        \loop
        \ifnum  \count 0 < #1
        \then   \advance \count 0 by 1
            \Mess@ge {Iteration \the \count 0 \space}%
            \Multiply \dimen 0 by {\dimen 2}%
            \Mess@ge {After multiplication, term = \nodimen 0}%
            \Divide \dimen 0 by {\count 0}%
            \Mess@ge {After division, term = \nodimen 0}%
        \repeat
        \Mess@ge {Final value for term #1 of
                \nodimen 2 \space is \nodimen 0}%
        \xdef \Term {#3 = \nodimen 0 \r@dians}%
        \aftergroup \Term
           }}
    \catcode `\p = \other
    \catcode `\t = \other
    \gdef \n@dimen #1pt{#1} 
}

\def \Divide #1by #2{\divide #1 by #2} 

\def \Multiply #1by #2
       {{
    \count 0 = #1\relax
    \count 2 = #2\relax
    \count 4 = 65536
    \Mess@ge {Before scaling, count 0 = \the \count 0 \space and
            count 2 = \the \count 2}%
    \ifnum  \count 0 > 32767 
    \then   \divide \count 0 by 4
        \divide \count 4 by 4
    \else   \ifnum  \count 0 < -32767
        \then   \divide \count 0 by 4
            \divide \count 4 by 4
        \else
        \fi
    \fi
    \ifnum  \count 2 > 32767 
    \then   \divide \count 2 by 4
        \divide \count 4 by 4
    \else   \ifnum  \count 2 < -32767
        \then   \divide \count 2 by 4
            \divide \count 4 by 4
        \else
        \fi
    \fi
    \multiply \count 0 by \count 2
    \divide \count 0 by \count 4
    \xdef \product {#1 = \the \count 0 \internal@nits}%
    \aftergroup \product
       }}

\def\r@duce{\ifdim\dimen0 > 90\r@dian \then   
        \multiply\dimen0 by -1
        \advance\dimen0 by 180\r@dian
        \r@duce
        \else \ifdim\dimen0 < -90\r@dian \then  
        \advance\dimen0 by 360\r@dian
        \r@duce
        \fi
        \fi}

\def\Sine#1%
       {{%
    \dimen 0 = #1 \r@dian
    \r@duce
    \ifdim\dimen0 = -90\r@dian \then
       \dimen4 = -1\r@dian
       \c@mputefalse
    \fi
    \ifdim\dimen0 = 90\r@dian \then
       \dimen4 = 1\r@dian
       \c@mputefalse
    \fi
    \ifdim\dimen0 = 0\r@dian \then
       \dimen4 = 0\r@dian
       \c@mputefalse
    \fi
    \ifc@mpute \then
        \divide\dimen0 by 180
        \dimen0=3.141592654\dimen0
        \dimen 2 = 3.1415926535897963\r@dian 
        \divide\dimen 2 by 2 
        \Mess@ge {Sin: calculating Sin of \nodimen 0}%
        \count 0 = 1 
        \dimen 2 = 1 \r@dian 
        \dimen 4 = 0 \r@dian 
        \loop
            \ifnum  \dimen 2 = 0 
            \then   \stillc@nvergingfalse
            \else   \stillc@nvergingtrue
            \fi
            \ifstillc@nverging 
            \then   \term {\count 0} {\dimen 0} {\dimen 2}%
                \advance \count 0 by 2
                \count 2 = \count 0
                \divide \count 2 by 2
                \ifodd  \count 2 
                \then   \advance \dimen 4 by \dimen 2
                \else   \advance \dimen 4 by -\dimen 2
                \fi
        \repeat
    \fi
            \xdef \sine {\nodimen 4}%
       }}

\def\Cosine#1{\ifx\sine\UnDefined\edef\Savesine{\relax}\else
                     \edef\Savesine{\sine}\fi
    {\dimen0=#1\r@dian\advance\dimen0 by 90\r@dian
     \Sine{\nodimen 0}
     \xdef\cosine{\sine}
     \xdef\sine{\Savesine}}}

\def\psdraft{
    \def\@psdraft{0}
}
\def\psfull{
    \def\@psdraft{100}
}

\psfull

\newif\if@scalefirst
\def\psscalefirst{\@scalefirsttrue}
\def\psrotatefirst{\@scalefirstfalse}
\psrotatefirst

\newif\if@draftbox
\def\psnodraftbox{
    \@draftboxfalse
}
\def\psdraftbox{
    \@draftboxtrue
} \@draftboxtrue

\newif\if@prologfile
\newif\if@postlogfile
\def\pssilent{
    \@noisyfalse
}
\def\psnoisy{
    \@noisytrue
} \psnoisy
\newif\if@bbllx
\newif\if@bblly
\newif\if@bburx
\newif\if@bbury
\newif\if@height
\newif\if@width
\newif\if@rheight
\newif\if@rwidth
\newif\if@angle
\newif\if@clip
\newif\if@verbose
\def\@p@@sclip#1{\@cliptrue}

\newif\if@decmpr

\def\@p@@sfigure#1{\def\@p@sfile{null}\def\@p@sbbfile{null}
            \openin1=#1.bb
        \ifeof1\closein1
                \openin1=\figurepath#1.bb
            \ifeof1\closein1
                    \openin1=#1
                \ifeof1\closein1%
                       \openin1=\figurepath#1
                    \ifeof1
                       \ps@typeout{Error, File #1 not found}
                        \if@bbllx\if@bblly
                        \if@bburx\if@bbury
                                \def\@p@sfile{#1}%
                                \def\@p@sbbfile{#1}%
                            \@decmprfalse
                        \fi\fi\fi\fi
                    \else\closein1
                            \def\@p@sfile{\figurepath#1}%
                            \def\@p@sbbfile{\figurepath#1}%
                        \@decmprfalse
                                \fi%
                \else\closein1%
                    \def\@p@sfile{#1}
                    \def\@p@sbbfile{#1}
                    \@decmprfalse
                \fi
            \else
                \def\@p@sfile{\figurepath#1}
                \def\@p@sbbfile{\figurepath#1.bb}
                \@decmprtrue
            \fi
        \else
            \def\@p@sfile{#1}
            \def\@p@sbbfile{#1.bb}
            \@decmprtrue
        \fi}

\def\@p@@sfile#1{\@p@@sfigure{#1}}

\def\@p@@sbbllx#1{
        \@bbllxtrue
        \dimen100=#1
        \edef\@p@sbbllx{\number\dimen100}
}
\def\@p@@sbblly#1{
        \@bbllytrue
        \dimen100=#1
        \edef\@p@sbblly{\number\dimen100}
}
\def\@p@@sbburx#1{
        \@bburxtrue
        \dimen100=#1
        \edef\@p@sbburx{\number\dimen100}
}
\def\@p@@sbbury#1{
        \@bburytrue
        \dimen100=#1
        \edef\@p@sbbury{\number\dimen100}
}
\def\@p@@sheight#1{
        \@heighttrue
        \dimen100=#1
        \edef\@p@sheight{\number\dimen100}
}
\def\@p@@swidth#1{
        \@widthtrue
        \dimen100=#1
        \edef\@p@swidth{\number\dimen100}
}
\def\@p@@srheight#1{
        \@rheighttrue
        \dimen100=#1
        \edef\@p@srheight{\number\dimen100}
}
\def\@p@@srwidth#1{
        \@rwidthtrue
        \dimen100=#1
        \edef\@p@srwidth{\number\dimen100}
}
\def\@p@@sangle#1{
        \@angletrue
        \edef\@p@sangle{#1} 
}
\def\@p@@ssilent#1{
        \@verbosefalse
}
\def\@p@@sprolog#1{\@prologfiletrue\def\@prologfileval{#1}}
\def\@p@@spostlog#1{\@postlogfiletrue\def\@postlogfileval{#1}}
\def\@cs@name#1{\csname #1\endcsname}
\def\@setparms#1=#2,{\@cs@name{@p@@s#1}{#2}}
\def\ps@init@parms{
        \@bbllxfalse \@bbllyfalse
        \@bburxfalse \@bburyfalse
        \@heightfalse \@widthfalse
        \@rheightfalse \@rwidthfalse
        \def\@p@sbbllx{}\def\@p@sbblly{}
        \def\@p@sbburx{}\def\@p@sbbury{}
        \def\@p@sheight{}\def\@p@swidth{}
        \def\@p@srheight{}\def\@p@srwidth{}
        \def\@p@sangle{0}
        \def\@p@sfile{} \def\@p@sbbfile{}
        \def\@p@scost{10}
        \def\@sc{}
        \@prologfilefalse
        \@postlogfilefalse
        \@clipfalse
        \if@noisy
            \@verbosetrue
        \else
            \@verbosefalse
        \fi
}
\def\parse@ps@parms#1{
        \@psdo\@psfiga:=#1\do
           {\expandafter\@setparms\@psfiga,}}
\newif\ifno@bb
\def\bb@missing{
    \if@verbose{
        \ps@typeout{psfig: searching \@p@sbbfile \space  for bounding box}
    }\fi
    \no@bbtrue
    \epsf@getbb{\@p@sbbfile}
        \ifno@bb \else \bb@cull\epsf@llx\epsf@lly\epsf@urx\epsf@ury\fi
}
\def\bb@cull#1#2#3#4{
    \dimen100=#1 bp\edef\@p@sbbllx{\number\dimen100}
    \dimen100=#2 bp\edef\@p@sbblly{\number\dimen100}
    \dimen100=#3 bp\edef\@p@sbburx{\number\dimen100}
    \dimen100=#4 bp\edef\@p@sbbury{\number\dimen100}
    \no@bbfalse
}
\newdimen\p@intvaluex
\newdimen\p@intvaluey
\def\rotate@#1#2{{\dimen0=#1 sp\dimen1=#2 sp
          \global\p@intvaluex=\cosine\dimen0
          \dimen3=\sine\dimen1
          \global\advance\p@intvaluex by -\dimen3
          \global\p@intvaluey=\sine\dimen0
          \dimen3=\cosine\dimen1
          \global\advance\p@intvaluey by \dimen3
          }}
\def\compute@bb{
        \no@bbfalse
        \if@bbllx \else \no@bbtrue \fi
        \if@bblly \else \no@bbtrue \fi
        \if@bburx \else \no@bbtrue \fi
        \if@bbury \else \no@bbtrue \fi
        \ifno@bb \bb@missing \fi
        \ifno@bb \ps@typeout{FATAL ERROR: no bb supplied or found}
            \no-bb-error
        \fi
        \count203=\@p@sbburx
        \count204=\@p@sbbury
        \advance\count203 by -\@p@sbbllx
        \advance\count204 by -\@p@sbblly
        \edef\ps@bbw{\number\count203}
        \edef\ps@bbh{\number\count204}
        \if@angle
            \Sine{\@p@sangle}\Cosine{\@p@sangle}
                {\dimen100=\maxdimen\xdef\r@p@sbbllx{\number\dimen100}
                        \xdef\r@p@sbblly{\number\dimen100}
                                \xdef\r@p@sbburx{-\number\dimen100}
                        \xdef\r@p@sbbury{-\number\dimen100}}
                        \def\minmaxtest{
               \ifnum\number\p@intvaluex<\r@p@sbbllx
                  \xdef\r@p@sbbllx{\number\p@intvaluex}\fi
               \ifnum\number\p@intvaluex>\r@p@sbburx
                  \xdef\r@p@sbburx{\number\p@intvaluex}\fi
               \ifnum\number\p@intvaluey<\r@p@sbblly
                  \xdef\r@p@sbblly{\number\p@intvaluey}\fi
               \ifnum\number\p@intvaluey>\r@p@sbbury
                  \xdef\r@p@sbbury{\number\p@intvaluey}\fi
               }
            \rotate@{\@p@sbbllx}{\@p@sbblly}
            \minmaxtest
            \rotate@{\@p@sbbllx}{\@p@sbbury}
            \minmaxtest
            \rotate@{\@p@sbburx}{\@p@sbblly}
            \minmaxtest
            \rotate@{\@p@sbburx}{\@p@sbbury}
            \minmaxtest
            \edef\@p@sbbllx{\r@p@sbbllx}\edef\@p@sbblly{\r@p@sbblly}
            \edef\@p@sbburx{\r@p@sbburx}\edef\@p@sbbury{\r@p@sbbury}
        \fi
        \count203=\@p@sbburx
        \count204=\@p@sbbury
        \advance\count203 by -\@p@sbbllx
        \advance\count204 by -\@p@sbblly
        \edef\@bbw{\number\count203}
        \edef\@bbh{\number\count204}
}
\def\in@hundreds#1#2#3{\count240=#2 \count241=#3
             \count100=\count240    
             \divide\count100 by \count241
             \count101=\count100
             \multiply\count101 by \count241
             \advance\count240 by -\count101
             \multiply\count240 by 10
             \count101=\count240    
             \divide\count101 by \count241
             \count102=\count101
             \multiply\count102 by \count241
             \advance\count240 by -\count102
             \multiply\count240 by 10
             \count102=\count240    
             \divide\count102 by \count241
             \count200=#1\count205=0
             \count201=\count200
            \multiply\count201 by \count100
            \advance\count205 by \count201
             \count201=\count200
            \divide\count201 by 10
            \multiply\count201 by \count101
            \advance\count205 by \count201
             \count201=\count200
            \divide\count201 by 100
            \multiply\count201 by \count102
            \advance\count205 by \count201
             \edef\@result{\number\count205}
}
\def\compute@wfromh{
        \in@hundreds{\@p@sheight}{\@bbw}{\@bbh}
        \edef\@p@swidth{\@result}
}
\def\compute@hfromw{
            \in@hundreds{\@p@swidth}{\@bbh}{\@bbw}
        \edef\@p@sheight{\@result}
}
\def\compute@handw{
        \if@height
            \if@width
            \else
                \compute@wfromh
            \fi
        \else
            \if@width
                \compute@hfromw
            \else
                \edef\@p@sheight{\@bbh}
                \edef\@p@swidth{\@bbw}
            \fi
        \fi
}
\def\compute@resv{
        \if@rheight \else \edef\@p@srheight{\@p@sheight} \fi
        \if@rwidth \else \edef\@p@srwidth{\@p@swidth} \fi
}
\def\compute@sizes{
    \compute@bb
    \if@scalefirst\if@angle
    \if@width
       \in@hundreds{\@p@swidth}{\@bbw}{\ps@bbw}
       \edef\@p@swidth{\@result}
    \fi
    \if@height
       \in@hundreds{\@p@sheight}{\@bbh}{\ps@bbh}
       \edef\@p@sheight{\@result}
    \fi
    \fi\fi
    \compute@handw
    \compute@resv}

\def\psfig#1{\vbox {
    \ps@init@parms
    \parse@ps@parms{#1}
    \compute@sizes
    \ifnum\@p@scost<\@psdraft{
        \special{ps::[begin]    \@p@swidth \space \@p@sheight \space
                \@p@sbbllx \space \@p@sbblly \space
                \@p@sbburx \space \@p@sbbury \space
                startTexFig \space }
        \if@angle
            \special {ps:: \@p@sangle \space rotate \space}
        \fi
        \if@clip{
            \if@verbose{
                \ps@typeout{(clip)}
            }\fi
            \special{ps:: doclip \space }
        }\fi
        \if@prologfile
            \special{ps: plotfile \@prologfileval \space } \fi
        \if@decmpr{
            \if@verbose{
                \ps@typeout{psfig: including \@p@sfile.Z \space }
            }\fi
            \special{ps: plotfile "`zcat \@p@sfile.Z" \space }
        }\else{
            \if@verbose{
                \ps@typeout{psfig: including \@p@sfile \space }
            }\fi
            \special{ps: plotfile \@p@sfile \space }
        }\fi
        \if@postlogfile
            \special{ps: plotfile \@postlogfileval \space } \fi
        \special{ps::[end] endTexFig \space }
        \vbox to \@p@srheight sp{
            \hbox to \@p@srwidth sp{
                \hss
            }
        \vss
        }
    }\else{
        \if@draftbox{
            \hbox{\frame{\vbox to \@p@srheight sp{
            \vss
            \hbox to \@p@srwidth sp{ \hss \@p@sfile \hss }
            \vss
            }}}
        }\else{
            \vbox to \@p@srheight sp{
            \vss
            \hbox to \@p@srwidth sp{\hss}
            \vss
            }
        }\fi

    }\fi
}} \psfigRestoreAt
\let\@=\LaTeXAtSign

\hfuzz=10pt \sloppy
\topmargin=-0.5cm
\renewcommand{\baselinestretch}{1.4}
\hfuzz=10pt \sloppy \oddsidemargin=1.0cm
 \textheight 212mm \textwidth=15.3cm
   \renewcommand{\theequation}{\arabic{equation}}

\newtheorem{definition}{Definition}[section]
\newtheorem{remark}{Remark}[section]

\def\u{u}
\def\e{\varepsilon}
\def\tt{\theta}
\def\defi{\stackrel{{\scriptscriptstyle \Delta}}{=}}
\def\bs{{\scriptscriptstyle BS}}
\def\OO{{\scriptscriptstyle O}}
\def\NO{{\scriptscriptstyle NO}}
\def\MM{{\scriptscriptstyle M}}
\def\PC{{\scriptscriptstyle PC}}
 \def\Cn{C([0,T];\R^n)}
\def\l{l}
\def\N{\mu}
\def\a{\alpha}
\def\d{\delta}
\def\o{\omega}
\def\O{\Omega}
\def\q{q}
\def\Y{{\cal Y}}
\def\F{{\cal F}}
\def\w{\widehat}
\def\Ind{{\mathbb{I}}}
\def\col{{\rm col\,}}
\def\sign{{\rm  sign\,}}
\def\mes{{\rm mes\,}}
\def\arctg{{\rm arctg\,}}
\def\diag{{\rm diag\,}}
\def\argmin{\mathop{\rm arg\, min}\,}
\def\argmax{\mathop{\rm arg\, max}\,}
\def\esssup{\mathop{\rm ess\, sup}}
\def\const{{\rm const\,}}
\def\es{{\rm ess\,}}
\def\Var{{\rm Var\,}}
\def\Tr{{\rm tr\,}}
\def\dist{{\rm dist\,}}
\def\R{{\bf R}}
\def\E{{\bf E}}
\def\P{{\bf P}}
\def\n{{\bf n}}
\def\S{{\bf S}}
\def\Z{{\cal Z}}
\def\PP{{\cal P}}
\def\J{{\cal J}}
\def\H{{\cal H}}
\def\ttt{{\widehat{\bf  t}}}
\def\h{h}
\def\L{L}
\def\z{{\zeta}}
\def\b{\beta}
\def\s{\delta}
\def\g{\gamma}
\def\C{{\bf C}}
\def\W{{\cal W}^*}
\def\W{{\cal W}}
\def\ww{\widetilde}
\def\X{{\cal X}}
\def\t{\theta}
\def\oo{\bar}
\def\s{\sigma}
\def\D{{\Delta}}
\def\p{\partial}
\def\G{\Gamma}
\def\GG{{\cal G}}
\def\U{{\cal U}}
\def\V{{\cal V}}
\def\A{{\cal A}}
\def\M{{\cal M}}
\def\B{{\cal B}}
\def\L{{\cal L}}
\def\I{{\, \cal I}}
\def\r{\rho}
\def\ThetaT{\Theta}
\def\h{h}
\def\RO{\stackrel{\circ}{\R}_+}
\def\ROn{\stackrel{\circ\ \ }{\R^n_+}}
\def\LLL{\w{\cal L}}
\def\LL{{\oo{\bf L}}}
\def\oneh{\frac{1}{2}}
\newcommand{\be}{\begin{equation}}
\newcommand{\ee}{\end{equation}}
\newcommand{\bd}{\begin{displaymath}}
\newcommand{\ed}{\end{displaymath}}
\newcommand{\ba}{\begin{array}{ll}}
\newcommand{\ea}{\end{array}}
\newcommand{\baa}{\begin{eqnarray}}
\newcommand{\eaa}{\end{eqnarray}}
\newcommand{\baaa}{\begin{eqnarray*}}
\newcommand{\eaaa}{\end{eqnarray*}}
\font\sm=cmr10
\def\BB{{\rm B}}
\def\KE{{\scriptscriptstyle KE}}
\def\PP{{\cal P}}
\def\T{{\cal T}}
\def\oopi{\oo\pi}
\def\wt{\hat}
\def\oo{\bar}
\def\nuu{\lambda}
\def\Fo{\F}
\def\BC{P_0}
\def\a{\alpha}
\def\iindex{******\index} 
\def\iindex{}
\title{
On detecting the dependence of time series\footnote{Accepted to
"Communications in Statistics -- Theory and Methods"; in press.
Submitted: 17 May 2010. Revised: 21 September 2010}}
\author{
Nikolai Dokuchaev\\ \ {\sm Department of Mathematics \& Statistics,
Curtin University,}\\
{\sm  GPO Box U1987, Perth, 6845 Western Australia}\\ {\sm email
N.Dokuchaev@curtin.edu.au}}

\begin{document}
 \maketitle
\begin{abstract}
This short note suggests a heuristic method for detecting  the
dependence of random time series that can be used in the case when
this dependence is relatively weak and such that the traditional
methods are not effective. The method requires  to compare some
special functionals on the sample characteristic functions with the
same functionals computed for the benchmark time series with a known
degree of
correlation. Some experiments for financial time series are presented.  \\
{\bf Key words}:  serial dependence,  non-parametric
  methods, technical analysis, econometrics
\end{abstract}
This short note presents some statistical experiments with the
purpose to estimate  the dependence for  time series.
 We  suggest to compare historical time series with a
series with given and known correlation using a functional formed
from empirical characteristic functions defined similarly to Hong
(1999). It gives a simple empirical method that allows to estimate
the dependence by comparing the values of this functional for two
time series.  The suggested test can be an addition to dependence
and correlation tests such as Pearson test, Hoeffding's test,
Spearman test, Kendall Tau Rank  test, or chi-square test; see,
e.g., Conover (1999) and Hollander and Wolfe (1999). In our
experiments, we used a simple autoregression as the benchmark
process. We present some results of experiments for  time series
with  admittedly weak dependence such as financial time series for
returns of stock prices.
\par
Note that the problem of detecting the serial correlations for
financial series is very important for applications in finance. In
particular, this problem is related to the open  problem of
validation of "technical analysis" methods that offer trading
strategies based on historical observations. The main benefit is
that these strategies are model-free: they require only historical
data. This is why they are so popular among traders. There are many
different strategies suggested in the framework of "technical
analysis".
  Hsu and Kuan (2005) mentioned that there are more
than 18,326 different empirical trading rules
 being used in practice. However, the question remains open if the
   main hypothesis of technical analysis is correct.  This  hypothesis
   suggests that it is possible
   to make a statistically reliable forecast for future stock price movements
   using recent prices, and, finally,
to find "winning" in statistical sense trading strategies. However,
the dependence from the past (if any) if extremely weak for the
stock prices, and this dependence is difficult to catch by usual
statistical methods. Statistical studies of historical prices
 made as early as in 1933 didn't support the hypothesis that there
 is
significant dependence from the past and predictability for the
stock prices; see
 the discussion and the bibliography in Chapter 2, pp. 37-38, from Shiryaev
 (1999). This is the reason why the most common and mainstream model for the stock prices is
 the random walk or its modifications.  Recently, new efforts were devoted to this problem, and some signs
 of possible presence of statistically significant   dependence from the past
 were found
 (see, e.g., Lo {\it et al.} (2000), Hsu and
Kuan (2005)),  Lorenzoni {\it et al} (2007)). In particular,
Lorenzoni {\it et al} (2007) found that, for a certain models of
stock price evolution, there is  a statistically significant
informational content in some patterns from technical analysis. The
computational experiments with our tests also show that the
financial time series have some dependence.

 The  paper is organized as follows. In Section 1 we describe the
 method and collect the
notation and definitions. Sections 2 contains description of the
experiments with financial time series. Sections 3 contains
conclusions and some suggestions for future research.
\section{The method}
\def\EE{{\mathbb{E}}}

Clearly, if random variables $\xi$ and $\eta$ are independent, then
\baa e(q)=|\EE e^{i \xi q} \EE e^{i \xi q} -\EE e^{i (\xi+\eta)
q}|\equiv 0,\quad q\in\R. \label{char} \eaa In (\ref{char}), $\EE$
denote the expectation, $i=\sqrt{-1}$ is the imaginary unit.
Condition (\ref{char}) is a necessary but  not a sufficient
condition of independence; see example in Hamedani and Volkmer
(2009). If (\ref{char}) holds then $\xi$ and $\eta$ are said to be
subindependent.
\par
 We
suggest to measure the sample analog of the function (\ref{char})
for the time series and their past history and match it with the
similar function for a benchmark AR(1) series with given correlation
coefficient.

Let $R_t$ be a time series (not necessary a stationary time series).
 Let
$G_t=\{R_k\}^{k=t}_{k=-\infty}=(R_t,R_{t-1},R_{t-2},....)$ represent
the history of the series. We are interested in detecting the
dependence of the current value of the series from the history.
\par
We denote by $\E$ the sample mean over available historical data.
Let $\ell_{\infty}$ denote the set of all bounded sequences
$\{x_k\}_{k=0}^{+\infty}\subset\R$.
\par
 We
suggest to calculate the values \baa e(h,F,q)=\Bigl|{\E
e^{iqh(G_{t-1})} \E e^{ iq F(R_t)}}-{\E
e^{iqh(G_{t-1})+iqF(R_t)}}\Bigr|, \label{eq}\eaa where
$h:\ell_{\infty}\to\R$ and $F:\R\to\R$  are some real valued
functions, $q\in\R$.

If $e(q)$ is sufficiently different from zero under some statistical
degree, then one is provided with empirical evidence in favor of
dependence.

To measure the degree of the dependence, we suggest to compare a
 norm of function (\ref{eq}) with the same norm  of a similar function
 calculated  for some benchmark the time series
$\{\ww R_t\}$ with certain given level of correlations.  We suggest
to use as the benchmark series the time series generates by
autoregression AR(1) \baa \ww R_t=a\ww R_{t-1}+\e_t,\label{ar}\eaa
where $a\in (-1,1)$, $\e_t$ are samples from independent identically
distributed random variables such that $\EE \e_t=0$. These series
can be created using Monte-Carlo simulation.
\par
 Let
$\ww G_t=\{\ww R_k\}^{k=t}_{k=-\infty}=(\ww R_t,\ww R_{t-1},\ww
R_{t-2},....)$ . Let $\ww e(h,F,q,a)$ be the corresponding value
(\ref{eq}) calculated with $(R_t,G_t)$ replaced by $(\ww R_t,\ww
G_t)$.
\par
The following definition   is rather  heuristic but still gives an
idea how to measure the dependence from the history.
\begin{definition}\label{def}
\begin{itemize}
\item[(i)] Let functions $h$ and $F$ be given. Let a sample $\{R_t\}$ be given.
and let AR(1) autoregression $\{\ww R_t\}$  be defined by (\ref{ar})
with some  given coefficient $a\in\R$. We say that the sets of
characteristics $(h(G_{t-1},F(R_t)))$ and $(h(\ww G_{t-1},F(\ww
R_t)))$ have the same level of dependence from the history given
$(h,F)$ if $e(h,F,q)$ is similar in some sense to $\w e(h,F,q,a)$.
\item[(ii)] Let a set $\cal P$ of the pairs of functions $(h,F)$ be given. We say that the sample
 $\{R_t\}$
 have the same level of dependence from the history
as autoregression   $\{\ww R_t\}$ defined by (\ref{ar}) with the
coefficient $\w a=\w a({\cal P})$ if this $|\w a|$  is the supremum
over all $|a|$ such that there exists $(h,F)\in{\cal P}$ such that
$(h(G_{t-1},F(R_t)))$ and $(h(\ww G_{t-1},F(\ww R_t)))$ have the
same level of dependence from the history.
\end{itemize}
\end{definition}
\begin{remark}{\rm
In Definition \ref{def}(i), the nature of the required similarity is
not specified. In the  experiments described below, we have assumed
that the similarity is achieved when the $\sup_{q\in[0,\oo q]}
e(h,F,q)=\sup_{q\in[0,\oo q]}\w e(h,F,a,q)$ for $\oo q>0$ defined by
computational abilities and decay of the functions; the interval
$[0,\oo q]$ should be large enough. In other words, we accepted that
the similarity is achieved  when these functions have the same norm
in $L_{\infty}(0,\oo q)$. So far, we have not compare this choice
with other possible choices such as comparison of integrals
$\int_0^{\oo q} e(h,F,q))dq$ and $\int_0^{\oo q} e(h,F,q))dq$.
}\end{remark}
 Note that it follows from the definitions that
$e(h,F,q)\equiv e(h,F,-q)$ and $\w e(h,F,a,q)\equiv\w e(h,F,a,q)$;
therefore, it suffices to consider  $q\ge 0$.
\section{Statistical experiments}
 We have carried out the following
 experiment for the time series representing the  returns for the historical stock prices, i.e., when
 $R_t\defi S_t/S_{t-1}-1$  where $S_t$ are the stock prices.
 Using
daily price data from 1984 to 2009 for 19 American and Australian
stocks (Citibank, Coca Cola, IBM, AMC, ANZ, LEI, LLC, LLN, MAY, MLG,
MMF, MWB, MIM, NAB, NBH, NCM, NCP, NFM and NPC), we generated
samples of price data for one synthetic return as
$\{R_t\}\defi\{S_t/S_{t-1}-1\}$, where $S_t$ is the price at day
$t$.  In fact,  the full 47 years of data was not available for all
the stocks; we have the size of sample equal to 69,948.  Since a
conclusion about a technical analysis strategy can only be made
after one collects the results of using it as many times as possible
(i.e., either for different stocks or for different time intervals),
we claim that our model and our experiment are not unreasonable.
\par
Let $\Ind$ denote the indicator function.
\par
\begin{figure}[ht]
\caption[]{Shapes for $e(h,F,q)$  for $(h,F)$ defined by (\ref{v1})
 and for $\w e(h,F,a,q)$ with $a=0.1$, $q\in [0,180)$, $\Var\e_t=(1-a^2)V$;
-----: values of $e(h,F,q)$; $-\, -\,- $: values of $\w e(h,F,a,q)$.}
\vspace{0.5cm} \centerline
{\psfig{figure=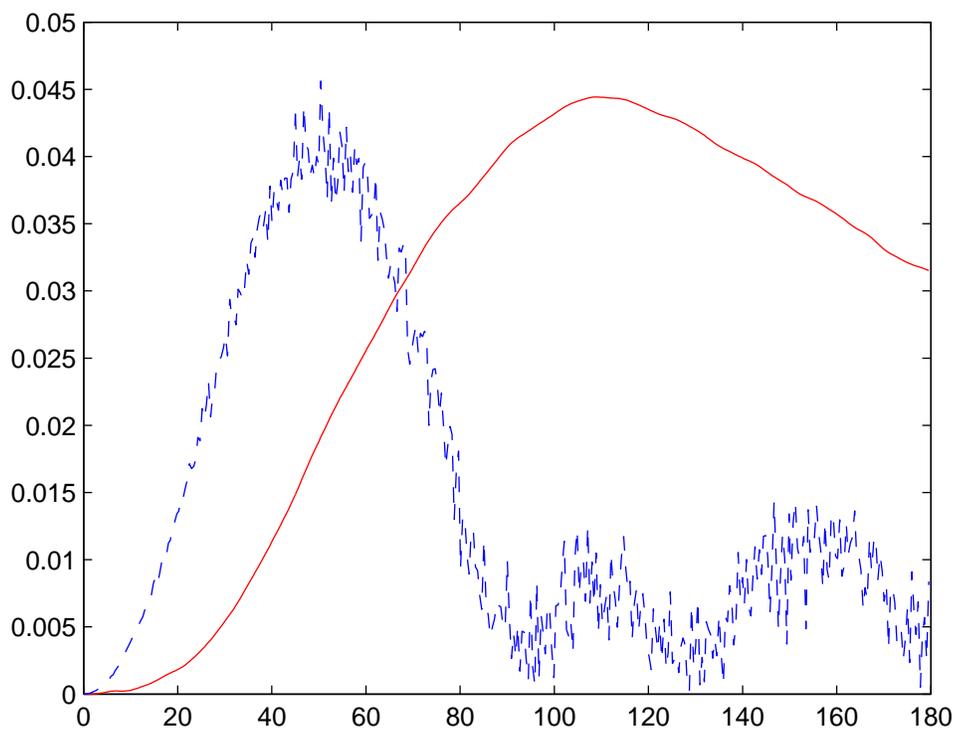,height=11.0cm}}
\label{fig1}
\end{figure}
\begin{figure}[ht]
\caption[]{Shapes for $e(h,F,q)$  for $(h,F)$ defined by (\ref{v2})
 and for $\w e(h,F,a,q)$ with $a=0.10$, $q\in [0,50)$;
-----: values of $e(h,F,q)$; $-\, -\,- $: values of $\w e(h,F,a,q)$.}
\vspace{0.5cm} \centerline
{\psfig{figure=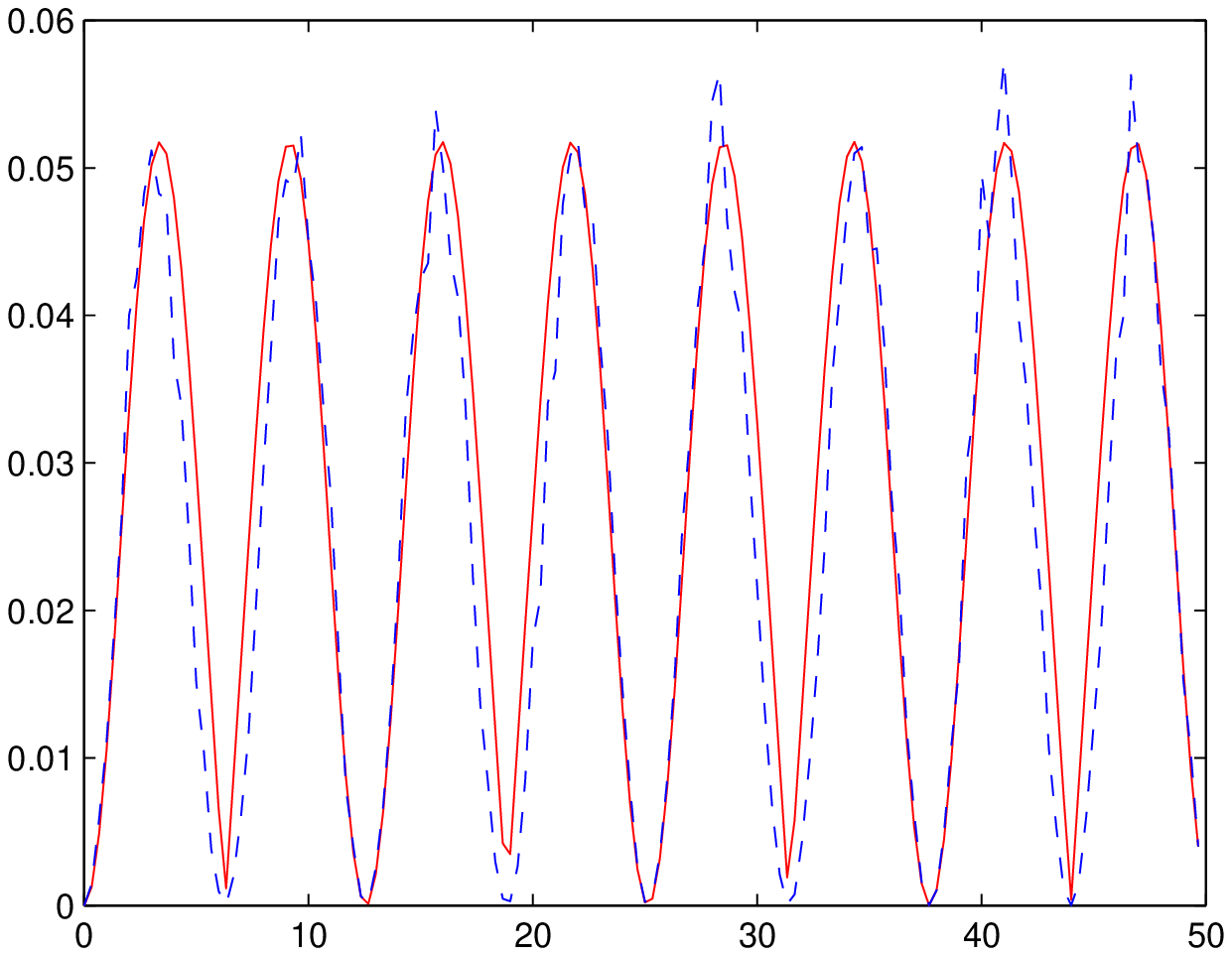,height=11.0cm}}
\label{fig2}
\end{figure}
\begin{figure}[ht]
\caption[]{Shapes for $e(h,F,q)$  for $(h,F)$ defined by (\ref{v3})
with $d=+\infty$
 and for $\w e(h,F,a,q)$ with $a=0.15$, $q\in [0,50)$;
-----: values of $e(h,F,q)$; $-\, -\,- $: values of $\w e(h,F,a,q)$.
} \vspace{0.5cm} \centerline
{\psfig{figure=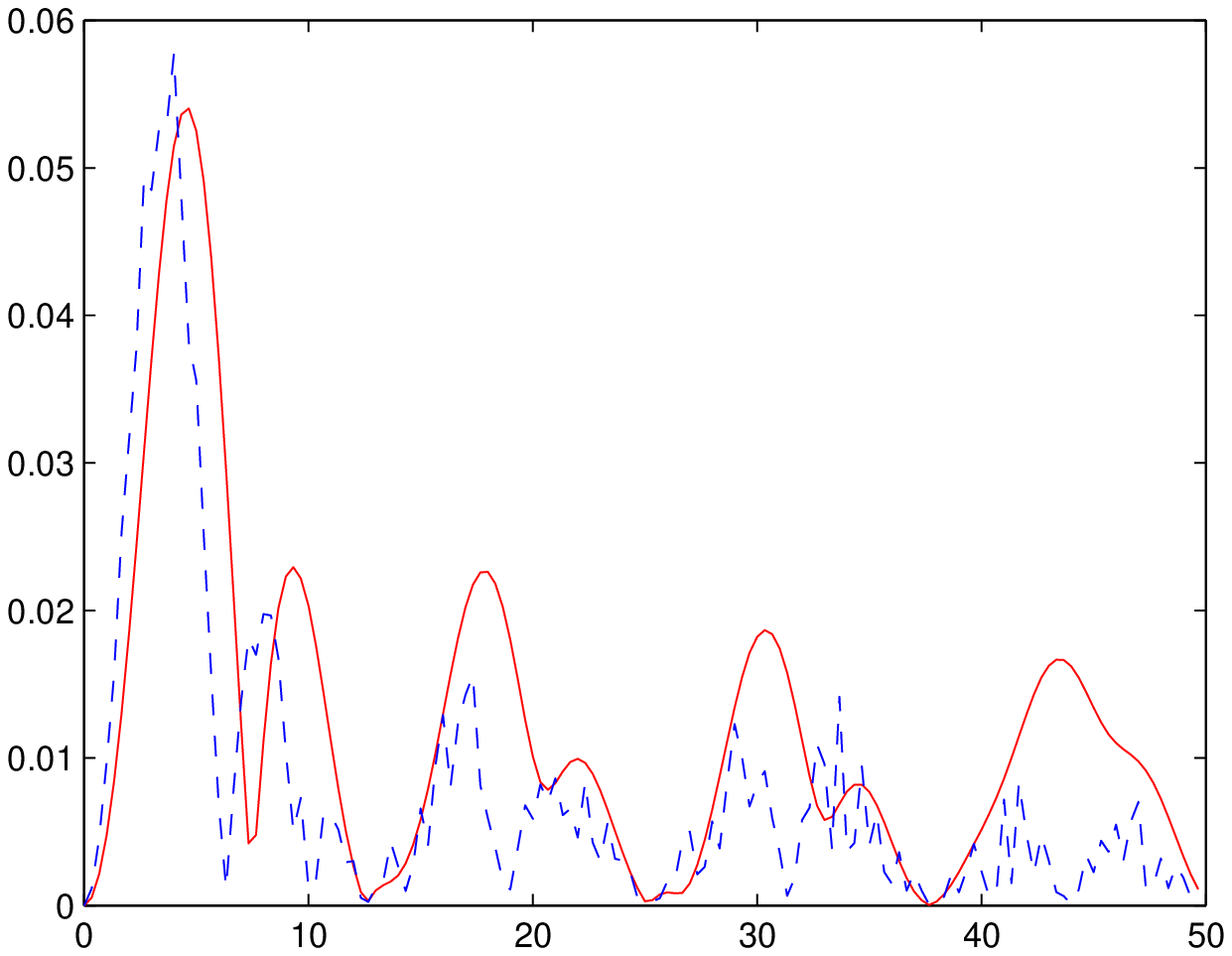,height=11.0cm}}
\label{fig3}
\end{figure}

 Let us describe the the analysis done for three possible
choices of functions $h$ and $F$.
\begin{itemize}
\item[]{\bf Choice 1:} \baa
h(G_{t-1})=R_{t-1},\quad F(R_t)=R_t. \label{v1} \eaa
 We found that $a(h,F)=0.1$ for
this case (see Fig.\ref{fig1}).
\item[]{\bf Choice 2:}
\baa
 h(G_{t-1})=\Ind_{\{ R_{t-1}>0\}},\quad F(R_t)=\Ind_{\{ R_{t-1}>0\}}.
\label{v2}
\eaa
 We found that $a(h,F)=0.1$ for
this case (see Fig.\ref{fig2}).
\item[]{\bf Choice 3.}
\baa &&h(G_{t-1}\})= \frac{1}{2}\Ind_{\{
R_{t-1}>0\}}+\frac{1}{2^2}\Ind_{\{
R_{t-2}>0\}}+....+\frac{1}{2^d}\Ind_{\{ R_{t-d}>0\}},
\nonumber\\
&&F(R_t)=\Ind_{\{ R_t>0\}}, \label{v3}
 \eaa
where $d$ is given.  We found that $a(h,F)=0.15$ for this case with
$d=+\infty$ (see Fig.\ref{fig3}).
\end{itemize}
\par
  Figures \ref{fig1}-\ref{fig3} show the samples of $e(h,F,q)$ and $\w
e(h,F,a,q)$ for $(h,F)$ defined by (\ref{v1})-(\ref{v3})
respectively, and for $a=a(h,H)$ that $max_qe(h,F,q)=\max_q\w
e(h,F)$.
\subsubsection*{On the choice of $(h,F)$} 
\begin{figure}[ht]
\caption[]{Shapes for $e(h,F,q)$  for $(h,F)$ defined by (\ref{v3})
with $d=+\infty$
 and for $\w e(h,F,a,q)$ with $a=-0.12$, $q\in [0,50)$;
-----: values of $e(h,F,q)$; $-\, -\,- $: values of $\w e(h,F,a,q)$.
} \vspace{0.5cm} \centerline
{\psfig{figure=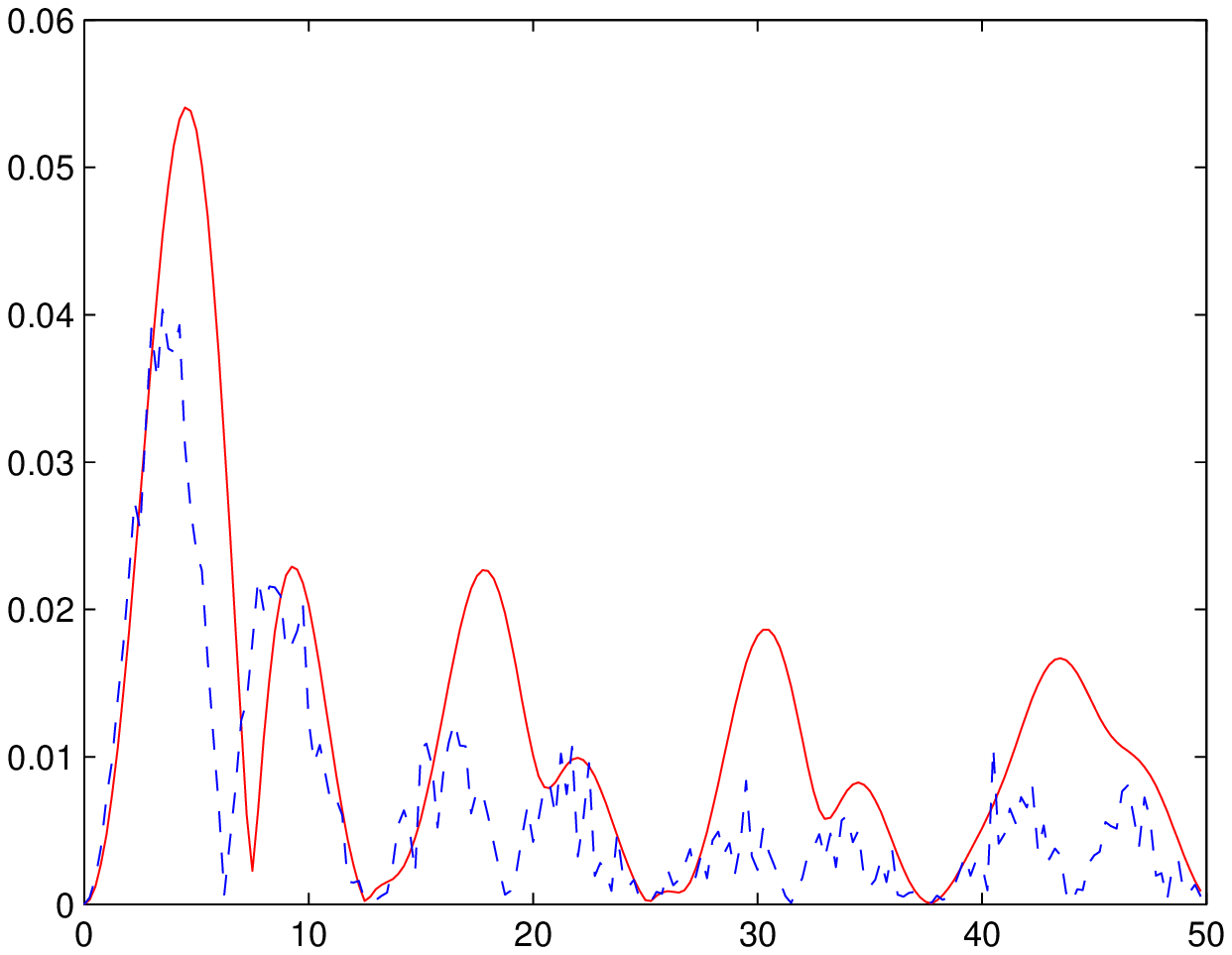,height=11.0cm}}
\label{fig4}\end{figure}

The functions $(h,F)$ in
Choices 1-3 were not selected by some optimal way; we leave it for
future research. However, there are certain reasons for the
particular Choices 2-3 for the functions $h$ and $F$ instead of more
straightforward Choice 3. First, the binary characterization of
increments helps to reduce calculations. Second, the Choices 2-3
reduces the impact of volatility and give more emphasize  on price
movement in the spirit of technical analysis for stock trading,
where the sign of the changes is crucial. Special selection of $h$
in the
 Choice 2 allows to take in account all the history of the signs of the price
movements (i.e., the history reduced to the binary characteristics);
the impact of older movements decays exponentially. In our
experiments, we found that Choices 2 and 3 ensures the most robust
results; the corresponding curves on the graphs generated by the
benchmark AR(1) series behave very regularly with respect to small
changes of $|a|$. It can be illustrated by Fig \ref{fig3} and Fig
\ref{fig4}, where the results for Choice 3 with $a=0.15$ and
$a=-0.12$ respectively are presented. The conditions of the
experiment were the same except the choice of $a$.

In some other experiments that we leaved outside of this paper, we
found that different combinations of $h$ and $F$ from Choices 1-3
also give robust results that are close to the results presented
here. (In particular, we used the pair consisting of function $F$
from Choice 1 and function $h$ from Choice 3).

The experiments show  that the maximum matching value of $a(h,F,q)$
can be achieved for the choice of $(h,F)$ defined by (\ref{v3}).
Moreover, this result is quite robust with respect to variations of
the parameters and data sets. The graphs are practically not
changing if we remove any subset from the set of 19 stocks.

For these experiments, we developed a simple MATLAB programm. This
programm cannot run over a set of $a$, so the graphs for every
particular $a$ were analyzed one by one. For every particular $a$,
this programm gives the answer for the question: {\em Is dependence
of underlying time series is stronger or weaker that the dependence
of AR(1) with the coefficient $a$?} Obviously, a better programm
could make automatic calculation of the best matching  $a$.
\subsubsection*{On the direct computing
the correlation of coefficient} In addition, we tested the
hypothesis that the series for $R_t$ is described as linear AR(1)
autoregression  \baaa R_t=\b R_{t-1}+\e_t. \eaaa The standard least
square estimator gives the value $\w\b=-0.005$ that is too small to
indicate the presence
 of correlation. The same value $\w\b=-0.005$ was obtained  from
 Pearson Product-Moment
Correlation test for $R_t$ and $R_{t-1}$. Moreover, this value
$\w\b$ coefficient appears to be non-robust with changes of the data
set; it varies significantly if we add or delete a particular stock.

On the other hand, we found, using our test, that the series $R_t$
has the same degree of dependence as AR(1) regression (\ref{ar})
with coefficient $a=0.15$. Therefore, we can conclude that the
dependence cannot be expressed via straightforward calculation of
the correlation as the coefficient for the linear autoregression
model.

\subsection*{Selection of $\sign a$ and $\Var \e_t$ for the benchmark
series}
\begin{remark}{\rm In our criterion, we use only the value $|a|$
and ignore the sign for the coefficient $a$ that defines the
correlation of the benchmark series. The reason is that, as we
observed in the experiments, the shape of $\w e(h,F,q,a)$ is very
close to the shape of $\w e(h,F,q,-a)$. It can be illustrated by Fig
\ref{fig3} and Fig \ref{fig4}, where the results for Choice 3 with
$a=0.15$ and $a=-0.12$ respectively are presented. The conditions of
the experiment were the same except the selection of $a$. Other
experiments showed the same independence from the sign of $a$ for
other choices of $(h,F)$. }\end{remark}
\par
In  the experiments, we considered benchmark series with $\w R_0=0$.
For the Choices 2 and 3, the results are not affected by the
selection of the value for $\Var \e_t$. We used $\Var\e_t=1$ for the
Choices 2 and 3. For the Choice 1, the selection of $\Var \e_t$
defines the scaling of the function $\w e(h,F,a,q)$: for instance,
let the series $\e_t$ generates the function $\w e(h,F,a,q)$. If we
replace $\e_t$ by $k\e_t$ for some $k>0$, then the function $\w
e(h,F,a,q)$ will be replaced by the function $\w e(h,F,a,kq)$, i.e.,
the value of $\sup_q \e(h,F,q)$ will not be affected but the visual
image of the graph of the function will be changed.

We found that a convenient scaling can be achieved with
$\Var\e_t=(1-a^2)V$, where $V$ is the sample second moment for
$R_t$. In this case, $\Var\ww R_t$ is asymptotically close to $V$,
and it ensures a satisfactory scaling.

\section{Other choices of benchmark processes}
We have suggested to use the simplest  AR(1) series as the benchmark
series $\{R_t\}$. Alternatively,  other models with certain
predetermined level of serial dependence can be used, such as Markov
chains with a given size of non-diagonal elements in the matrix
transitional probabilities, or with an ARCH or GARCH process, or
with autoregression of a higher order.  These model with
multidimensional parameters have more flexibility. However, it is
more difficult to use them to order  the series $R_t$ with respect
to the degree of dependence.

Consider, for example, ARCH series for the purpose to generate a
benchmark dependence. Let us consider the following model for the
benchmark process: \baaa \ww R_{t+1}=a\ww R_t+\s_t\e_t,\qquad
\s_t=b+c\e_t^2. \eaaa For this model, the degree of the dependence
is defined by $(a,b,c)$. The case of $c=0$ corresponds to AR(1)
model that was used above. Matching the degree of dependence for the
ARCH model and for the observed series $\{R_k\}$ leads for situation
when the same degree of dependence can be achieved with  selection
of parameters $(a_1,b_2,c_1)$ and $(a_2,b_2,c_2)$, where
$(a_1,b_2,c_1)\neq (a_2,b_2,c_2)$. For example, we obtained that
 $(a,b,c)=(0.02,1,0.08\E \e_t^2)$ gives the same maximum of $\w e$ as
 $(a,b,c)=(0.1,0,0)$ (i.e.,
without ARCH), with $(h,F)$ defined by Choice 1. The corresponding
plot is shown on Fig. \ref{fig4.1} below. Therefore, the presence of
vector parameters  lead to analysis of one-dimensional surfaces
$\{(a,b,c)\}\in\R^3$ that correspond to different level of
dependence of stock returns. We leave it for future research.
\section{Conclusion} We  suggested a method that
allows to make a fast detecting of dependence from the past and some
estimate of the degree of dependence via comparison with a benchmark
AR(1) series. This method requires to  compare visually  the graphs
for functions $e$ and $\w e$. This estimate is not very precise;
however, it is quite robust with respect to variations of the
parameters and data sets. We used this method in statistical
experiments with stock prices. We found that the result of the
experiments support the hypothesis that there is certain dependence
for financial time series. It gives a reason in favor of an
existence of a statistically winning strategy based on observations
of recent prices (i.e., a winning "technical analysis" trading
strategy).

\subsubsection*{Acknowledgment} \iindex{This work was
supported by NSERC grant of Canada 341796-2008 to the author.}
\section*{References}
 $\hphantom{xx}$
Conover, W. J. (1999). Practical Nonparametric Statistics. 3rd
edition. Wiley.

G. G. Hamedani, H. W. Volkmer. (2009). Letter to the Editor. {\em
The American Statistician } {\bf 63} (3), 295-295

Hsu, P.-H., Kuan, C.-M. (2005). Reexaming the profitability of
technical analysis with data snooping checks. {\it Journal of
Financial Econometrics} {\bf 3}, iss. 4, 606-628.
\par

Hollander and Wolfe (1999). Non-parametric statistical method.
Wiley.

Hong, Y. (1999) Hypothesis Testing in Time Series via the Empirical
Characteristic Function: A Generalized Spectral Density Approach
{\it Journal of the American Statistical Association} {\bf 94}, No.
448, 1201--1220.
\par

 Lo, A.W., Mamaysky, H., and Wang, Jiang. (2000).
Foundation of technical analysis: computational algorithms,
statistical inference, and empirical implementation.  {\em Journal
of Finance} {\bf 55} (4), 1705-1765.
\par
 Lorenzoni, G., Pizzinga, A., Atherino, R.,
 Fernandes, C,, Freire, R.R.. (2007). On the Statistical Validation of
Technical Analysis. {\it Revista Brasileira de Finan\c{c}as}. Vol.
5, No. 1, pp. 3–28.
\par
Shiryaev, A.N. (1999) Essentials of Stochastic Finance. Facts,
Models, Theory. World Scientific Publishing Co., NJ, 1999.

\begin{figure}[ht]
\caption[]{Shapes for $e(h,F,q)$
 and for $\w e(h,F,a,q)$ with $(h,F)$ defined for Choice 1 and for ARCH
 benchmark model
with $a=0.02$, $b=1$, $c=0.08\E\e_t^2$;
-----: values of $e(h,F,q)$; $-\, -\,- $: values of $\w e(h,F,a,q)$.}
\vspace{0.5cm}
 \centerline{\psfig{figure=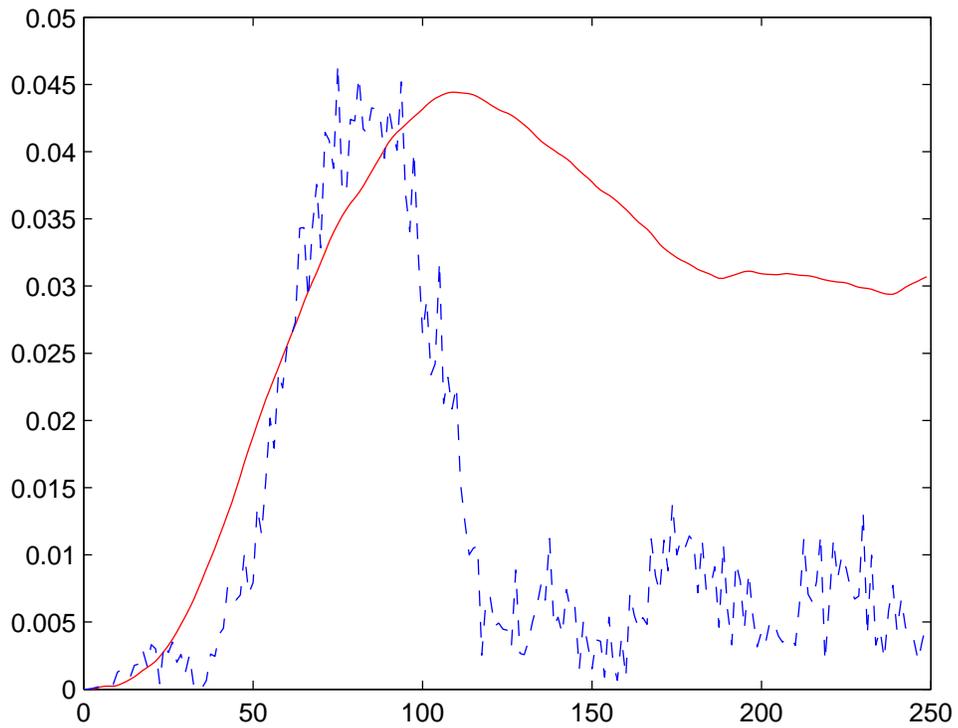,height=11.0cm}}
\label{fig4.1}
\end{figure}
\end{document}